# Enhancing Use Case Points Estimation Method Using Soft Computing Techniques

Ali Bou Nassif[*1], Luiz Fernando Capretz[2] and Danny Ho[3]

[*1]Electrical and Computer Engineering, University of Western Ontario, London, Ontario, Canada
Email: abounass@uwo.ca
[2]Electrical and Computer Engineering, University of Western Ontario, London, Ontario, Canada
Email: lcapretz@uwo.ca
[3]NFA Estimation Inc., Richmond Hill, Ontario, Canada
Email: danny@nfa-estimation.com

*Abstract*: Software estimation is a crucial task in software engineering. Software estimation encompasses cost, effort, schedule, and size. The importance of software estimation becomes critical in the early stages of the software life cycle when the details of software have not been revealed yet. Several commercial and non-commercial tools exist to estimate software in the early stages. Most software effort estimation methods require software size as one of the important metric inputs and consequently, software size estimation in the early stages becomes essential. One of the approaches that has been used for about two decades in the early size and effort estimation is called use case points. Use case points method relies on the use case diagram to estimate the size and effort of software projects. Although the use case points method has been widely used, it has some limitations that might adversely affect the accuracy of estimation. This paper presents some techniques using fuzzy logic and neural networks to improve the accuracy of the use case points method. Results showed that an improvement up to 22% can be obtained using the proposed approach.

## INTRODUCTION AND PROBLEM DEFINITION

As the role of software in the industry and the society becomes vital, it becomes crucial to develop high-quality and cost-effective software in a short period. To attain this goal, software development processes should be managed efficiently from the requirement phase to the implementation phase. One of the main tasks of project management is planning. Planning includes the cost and effort estimation of the project in the early stages of the software development life cycle. The earlier the estimation is, the better the project management will be. Even though early estimation is necessary, the accuracy of this estimation is very important. Software estimators are notorious with inaccurate estimation that leads to incomplete projects and consequently millions of dollars are wasted. The International Society of Parametric Analysis (ISPA) [1] and the Standish Group International [2] identified poor estimation as one of the main culprits behind software failure. Software cost and effort estimation mainly depend on the prediction of software size. This has led to the substantial increase in research in software engineering for estimating the size of software in the requirement stage.

Function Points Analysis (FPA) is one of the earliest models that is used to predict the size of software in the early stages. The FPA model was proposed by Albrecht in 1979 [3] and it measures the size of software based on its functionalities. The main advantages of the FPA model are that it is independent of the technology and the programming language used in the implementation. On the other hand, the main issues with the FPA model are that function points cannot be computed automatically and the decisions made in counting function points are subjective [4].

Object-Oriented Modelling (OOM) has become dominant since the release of Unified Modeling Language (UML) version 1.1 in 1997 [5], but OOM has become more popular since the release of UML 2.0 in 2005 [6]. UML models include use case, sequence, component, activity and class diagrams. Recently, many software organizations use UML notation to convey the requirements and the design of their software projects. For instance, use case, sequence and component diagrams might be used to represent the requirements of the system while the class diagram might be used to represent the system design.

One of the size and effort estimation models that rely on the use case diagram is called Use Case Points (UCP). The UCP model was proposed by Gustav Karner in 1993 [7]. UCP is measured by counting the number of use cases and the number of actors, each multiplied by its complexity factors. Use cases and actors are classified into three categories. These include *simple*, *average* and *complex*. The determination of the use cases' complexity (simple, average or complex) is determined by the number of transactions per use case. For instance, a use case is classified as *simple* if number of transactions is between one and three, classified as *average* if the number of transactions is between four and seven, classified as *complex* if the number of transactions is greater than seven.

The UCP presents some limitations that affect the accuracy of the estimation. The main drawback of this model is the absence of the graduation when classifying the complexity of the use cases. For example, if the number of the transactions in a use case is three, the use case is classified as *simple*, however, if the number of transactions is four, the use case is classified as *average*. According to the UCP, if project A contains ten use cases, each of three transactions and project B contains ten use cases, each of four transactions, then the size of project B will be double the size of project A. In practice, this approach is incorrect. Moreover, a use case of eight transactions has the same complexity factor as the use case of twenty transactions since this model does not distinguish between large, very large and super large use cases.

This paper introduces a new approach to overcome the limitations of the UCP. First, rather than classifying a use case as *simple*, *average*, or *complex*, the use case will be classified as $u_x$, such as $x \in [1,10]$ where x represents the number of transactions. This concludes that there will be ten degrees of complexity for use cases ($u_1$, $u_2$, $u_3$, etc.). The proposed approach will be implemented in two independent stages. First, a fuzzy logic approach is applied to determine the complexity factor of $u_x$. The second stage of the proposed approach is implemented through a neural network model. The neural network model is a black box that takes $u_x$ (10 vectors) as an input, in addition to three vectors which represent the three types of the actors (simple, average or complex). The output of the neural network will be the size of the software.

The rest of the paper is organized as follows: Section 2 presents the background and the related work for the proposed approach. Sections 3 and 4 propose the fuzzy logic and the neural network approaches respectively. Section 5 evaluates the proposed approaches. Section 6 presents general discussion about the paper. Section 7 highlights the threats to validity in this work. Finally, section 8 concludes the paper and proposes the future work.

## BACKGROUND AND RELATED WORK

This paper presents a new approach to improve the accuracy of the use case estimation model using fuzzy logic and neural network. This section presents the terms that are relevant to this work.

**Use Case Points**

This method is based on mapping a use case diagram to a size metric called use-case points. When the size of software is known, the software development effort can be estimated. The use case model was first proposed by Jacobson et al. [8] A use case diagram shows how users interact with the system. A use case diagram is composed of use cases and actors. Use cases represent the functional requirements where an actor is a role played by a user. In the use case diagram, a use case can extend or include another use case. Figure 1 is an example of a use case diagram [9].

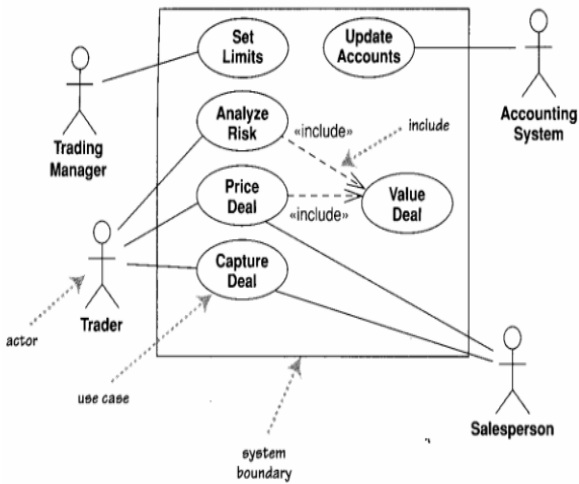

Figure 1: Use Case Diagram [9]

The use case points method mainly depends on four factors. These include the number and the complexity of the use cases, the number and the complexity of the actors, some non-functional requirements such as usability and portability, and some environmental factors where the software will be developed. The complexity of a use case is determined by the number of transactions of the use case scenario. A use case scenario is usually composed of several points. These include the actors involved in the scenario, the precondition of the system, the main success scenario, the extensions or exceptions and the post condition. The following example introduces the scenario of the use case "Student Enrolls in a Course" in a University Online Registration System.

**Use Case Title:** Student Enrolls in a Course
**Actors:** Student, Admin
**Precondition:** The student is not enrolled in a course
**Main Success Scenario:**
1. Check if the student has permission to register a course
2. Student chooses the course he or she wishes to enroll in
3. System checks for the deadline of enrollment
4. System checks for the prerequisite of the course
5. System checks if the student has registered in another course which is scheduled at the same time
6. System checks for the maximum number of courses the student can register
7. System checks if the course is full

**Extensions**
1a: The student does not have permission (e.g. the student has not paid the tuition)
    1a1: Notify the student to contact the administrator
3a: The deadline has passed
    3a1: An Error message will be displayed
    3a2: The student will be informed to contact the registrar
4a: The prerequisite of the course is not fulfilled
    4a1: The student will be advised to contact the professor to obtain permission
    4b1: If the student has permission from the professor, the student will be advised to contact the registrar to enroll him/her in the course
5a: Two courses have the same schedule
    5a1: The student is advised to choose either one
6a: The number of the enrolled courses has been exceeded
    6a1: An error message will be displayed
7a: The course is full
    7a1: An error message will be displayed

**Post condition:** The student has enrolled in a course

With respect to counting the transactions in the scenario, the transactions should be counted in the success scenario as well as in the extensions. For example, the number of transactions of the above scenario will be fifteen. This includes seven transactions in the success scenario and eight transactions in the extensions (1a1 + 3a1 + 3a2 + 4a1 + 4b1 + 5a1 + 6a1 + 7a1). For instance, counting the number of transactions can be subjective and one might count 3a1 and 3a2 as one transaction. We argue in section 6 that counting the transactions in the extensions the same way as counting the transactions in the success scenario might lead to overestimation. Thus, we believe that counting the transactions in the extension part should be performed in a different way.

*Unadjusted and Adjusted Use Case Points:* To estimate the size of software using this method, several rules should be applied. These rules include [7]

- Identify the complexity of each use case: The complexity is said to be *Simple* if the number of transactions within this use case is between one and three. The complexity is *Average* if the number of transactions is between four and seven. The complexity is *Complex* if the number of transactions is eight or more.
- Assign a weight factor for each level of complexity for use cases: This factor depends on the type of the project. Usually, if the complexity level is *Simple*, the factor given is five. If the complexity level is *Average*, the factor given is ten. If the complexity level is *Complex*, the factor given is fifteen.

- Identify the complexity of each actor: An actor is defined as *Simple* if it is System Interface. An actor is defined as *Average* if it is Interactive or Protocol-Driven Interface. The actor is defined as *Complex* if it is a Graphical Interface.
- Assign a weight factor for each level of complexity for actors: This is similar to the weight factors given to use cases. The weight factor is one for *Simple*, two for *Average* and three for *Complex*.
- Calculate the total use case weight factor (UseCase_WeightFactor): This is the sum of all *Simple* use cases multiplied by their weighting factor + the sum of all *Average* use cases multiplied by their weighting factor + the sum of all *Complex* use cases multiplied by their weighting factor.
- Calculate the total actor weight factor (Actor_WeightFactor): Apply the same rule as above to calculate the total actor weight factor.
- Calculate the Unadjusted Use Case Points (UUCP): UUCP = UseCase_WeightFactor + Actor_WeightFactor. The Unadjusted Use Case Points can be expressed as:

$$UUCP = \sum_{i=1}^{6} n_i * W_i \qquad (1)$$

where $n_i$ is the number of items of variety i and $W_i$ is the complexity weight.

At this point, the UUCP is calculated. Some cost estimation methods such as SEER-SEM takes the UUCP as an input of software size to calculate the cost and effort of software development. Karner [7] proposed an effort estimation method based on the Adjusted Use Case Points (UCP). The UCP is calculated by multiplying the UUCP by the technical and environmental factors. The technical factors contribute to the complexity of the system where the environmental factors contribute to the efficiency of the system. Depending on the technical and environmental factors, the UCP can be same as, smaller or larger than the UUCP. At most, the UCP can be larger or smaller than the UUCP by 30%. The technical and environmental factors can be classified in Table 1 and Table 2 respectively.

Table 1: Technical Factors [7]

| $F_i$ | Factors Contributing to Complexity | $W_i$ |
|---|---|---|
| $F_1$ | Distributed Systems | 2 |
| $F_2$ | Application performance objectives | 1 |
| $F_3$ | End user efficiency | 1 |
| $F_4$ | Complex internal processing | 1 |
| $F_5$ | Reusability | 1 |
| $F_6$ | Easy Installation | 0.5 |
| $F_7$ | Usability | 0.5 |
| $F_8$ | Portability | 2 |
| $F_9$ | Changeability | 1 |
| $F_{10}$ | Concurrency | 1 |
| $F_{11}$ | Special security features | 1 |
| $F_{12}$ | Provide direct access for third parties | 1 |
| $F_{13}$ | Special user training facilities | 1 |

Table 2: Environmental Factors [7]

| $F_i$ | Factors contributing to efficiency | $W_i$ |
|---|---|---|
| $F_1$ | Familiar with Objectory | 1.5 |
| $F_2$ | Part-time workers | -1 |
| $F_3$ | Analyst capability | 0.5 |
| $F_4$ | Application experience | 0.5 |
| $F_5$ | Object oriented experience | 1 |
| $F_6$ | Motivation | 1 |
| $F_7$ | Difficult programming language | -1 |
| $F_8$ | Stable requirements | 2 |

The Adjustment Use Case points (UCP) can be expressed as:

$$UCP = UUCP * TF * EF \qquad (2)$$

where TF is the Technical Factor and the EF is the environmental factor. TF is calculated as:

$$TF = C1 + C2 \sum_{i=1}^{13} F_i * W_i \qquad (3)$$

where $C1 = 0.6, C2 = 0.01$ and $F_i$ is a factor that takes values 0 or 1 or 2 or 3 or 4 or 5. The value 0 means irrelevant while the value 5 means essential. The value 3 means that the factor is not very essential, neither irrelevant. For instance, if all the factors have the value of 3, the TF will be 1. On the other hand, the environmental factor EF is calculated as:

$$EF = C1 + C2 \sum_{i=1}^{8} F_i * W_i \qquad (4)$$

where $C1 = 1.4, C2 = -0.03$ and $F_i$ is a factor which is equivalent to the $F_i$ of the technical factor (i.e between 0 and 5). If all the factors have the value of 3, then the EF will be 1.

After the size of software is calculated in UCP, the effort to develop this software can be estimated. According to Karner, the effort required to complete one UCP is twenty person hours.

**Fuzzy Logic**

Fuzzy logic is derived from the fuzzy set theory that was proposed by Lotfi Zadeh in 1965 [10]. As a contrary to the conventional binary (bivalent) logic that can only handle two values *True* or *False* (1 or 0), fuzzy logic can have a truth value which is ranged between 0 and 1. This means that in the binary logic, a member is completely belonged or not belonged to a certain set, however in the fuzzy logic, a member can partially belong to a certain set. Mathematically, a fuzzy set *A* is represented by a membership function as follows:

$$Fz[x \in A] = \mu_A(x): \mathbb{R} \to [0, 1] \quad (5)$$

Where $\mu_A$ is the degree of the membership of element $x$ in the fuzzy set $A$.

A fuzzy set is represented by a *membership function*. Each element will have a grade of membership that represents the degree to which a specific element belongs to the set. Membership functions include *Triangular*, *Trapezoidal* and *S-Shaped*. In fuzzy logic, *linguistic variables* are used to express a rule or fact. For example, "the temperature is thirty degrees" is expressed in fuzzy logic by "the temperature is low" or "the temperature is high" where the words *low* and *high* are linguistic variables. In fuzzy logic, the knowledge based is represented by if-then rules. For example, if the temperature is high, then turn on the fan. The fuzzy system is mainly composed of three parts. These include *Fuzzification*, *Fuzzy Rule Application* and *Defuzzification*. Fuzzification means applying fuzzy membership functions to inputs. Fuzzy Rule Application is to make inferences and associations among members in different groups. The third step in the fuzzy system is to defuzzify the inferences and associations and make a decision and provide an output that can be understood. In this paper, fuzzy logic will be used to calibrate the complexity weight of use cases.

### Artificial Neural Network

Artificial Neural Network (ANN) is a network composed of artificial neurons or nodes which emulate the biological neurons [11]. ANN can be trained to be used to approximate a non-linear function, to map an input to an output or to classify outputs. There are several algorithms available to train a neural network but this depends on the type and topology of the neural network. The most prominent topology of ANN is the feed-forward networks. In a feed-forward network, the information always flows in one direction (from input to output) and never goes backwards. An ANN is composed of nodes organized into layers and connected through weight elements. At each node, the weighted inputs are aggregated, thresholded and inputted to an activation function to generate an output of that node. Mathematically, this can be represented by:

$$y(t) = f\left[\sum_{i=1}^{n} w_i\, x_i - w_0\right] \quad (6)$$

Where $x_i$ are neuron inputs, $w_i$ are the weights and $f[.]$ is the activation function.

Feed-forward ANN layers are usually represented as *input*, *hidden* and *output* layers. If the hidden layer does not exist, then this type of the ANN is called *perceptron*. The perceptron is a linear classifier that maps an input to an output provided that the output falls under two categories. The perceptron can map an input to an output if the relationship between the input and output is linear. If the relationship between the input and output is not linear, multi-layer perceptron (MLP) can be used. A MLP contains at least one hidden layer. MLPs can be trained using the backpropagation algorithm. In this paper, a MLP is used and trained using the backpropagation algorithm.

### Evaluation Criteria

Several methods exist to compare cost estimation models. Each method has its advantages and disadvantages. In this work, three methods will be used. These include the Mean of the Magnitude of Relative Error (MMRE), the Mean of Magnitude of error Relative to the Estimate (MMER) and the Mean Error with Standard Deviation.

*MMRE:* This is a very common criterion used to evaluate software cost estimation models [12]. The Magnitude of Relative Error (MRE) for each observation i can be obtained as:

$$MRE_i = \frac{|\,Actual\ Effort_i - Predicted\ Effort_i\,|}{Actual\ Effort_i} \quad (7)$$

MMRE can be achieved through the summation of MRE over $N$ observations:

$$MMRE = \frac{1}{N}\sum_{1}^{N} MRE_i \quad (8)$$

*MMER:* MMER is another method for cost estimation models evaluation [13]. MER is similar to MRE with a difference that the denominator is the predicted effort instead of the actual effort. Consequently, the equations for MER and MMER are:

$$MER_i = \frac{|\,Actual\ Effort_i - Predicted\ Effort_i\,|}{Predicted\ Effort_i} \quad (9)$$

$$MMER = \frac{1}{N}\sum_{1}^{N} MER_i \quad (10)$$

When using the MMRE and the MMER in evaluation, good results are implied by lower values of MMRE and MMER.

*Mean Error with Standard Deviation:* Although MMRE and MMER have been used for a long time, both methods might lack accuracy. If the actual effort was small, MMRE would be high. On the other hand, if the predicted effort was low, MMER would also be high. Foss et al. argued that MMRE should not be used when comparing cost estimation models and using the standard deviation would be better [14]. The standard deviation method was first proposed by Karl Pearson in 1894 [15]. The equation for the mean error for each observation i and total number of observations N is:

$$\bar{x} = \frac{1}{N}\sum_{i=1}^{N} x_i \quad (11)$$

Where $x_i = (Actual\ Effort_i - Predicted\ Effort_i)$

The equation of the standard deviation can be seen as:

$$SD = \sqrt{\frac{1}{N-1}\sum_{i=1}^{N}(x_i - \bar{x})^2} \quad (12)$$

The mean error with standard deviation can be represented as:

$$\bar{x} \pm SD \qquad (13)$$

### Related Work

Little work has been done to improve the use case points model, however soft computing techniques such as fuzzy logic and neuro-fuzzy have been widely implemented to improve cost estimation models such as COCOMO II, Function Points Analysis and SEER-SEM. This section presents the work relevant to applying soft computing techniques on cost estimation models. These include the following:

Fetcke et al. [16] mapped UML use case diagrams to the software size metric Function Points. This method is based on four main steps. In the first step, Fetcke et al. define boundary concepts. This is similar to the boundary definition in FPA IFPUG. The authors suggest that actors are mapped into users and external applications, but the relationship is not always one-to-one. In the second step, the identification of items within the boundary is defined. In FPA, there are 2 types of items, *transactional functions* and *files* (data functions). Use cases are mapped in transactional functions. In order to count transactional functions, use cases must be described in further detail (use case scenarios). The concept of a file in Object Oriented is the object. The authors distinguish between typed objects and untyped objects. Each is treated in a specific way. Aggregation (Part-Of) and Inheritance (IS-A) relationships are also taken into consideration. In the third step, the identification of item types is defined. Transactional functions are treated as external outputs, external inquiries and external inputs. Files are treated as internal logical files and external interface files. The counting rules for transactional functions and files are the same as reported in the IFPUG Counting Practices Manual [17]. Finally, weight factors are applied. In this step, transactions and files are weighed based on IFPUG Counting Practices Manual.

Issa et al. [18], used the use case diagram of software to determine the effort of the software based on three steps. First, the effort estimation can be roughly calculated based on the number of use cases multiplied by 0.67 person-months. Secondly, estimation can be done using the use case patterns catalogue estimation method. Finally, object points can be extracted using the use case model method.

Mittal et al. [19], used fuzzy logic to tune the parameters of COCOMO cost estimation model. After that, a comparison between the proposed model and other models was conducted.

Huang et al. [20], proposed a new model using neuro-fuzzy technique to improve the estimation of the COCOMO model. This model can be easily trained and evaluated by experts. A learning algorithm for this model was also put forward.

### PROPOSED MODEL USING FUZZY LOGIC APPROACH

As explained in section 1, the main problem of the use case points model is that there is no graduation when classifying the complexity factors of use cases. In this section of the work, fuzzy logic with triangular membership was used to solve this issue. The input and output memberships are displayed in Figure 2 and Figure 3 respectively.

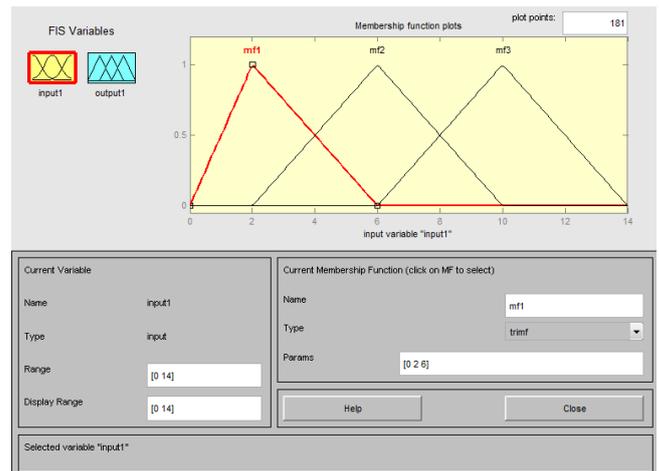

Figure 2: Fuzzy Logic Input Membership

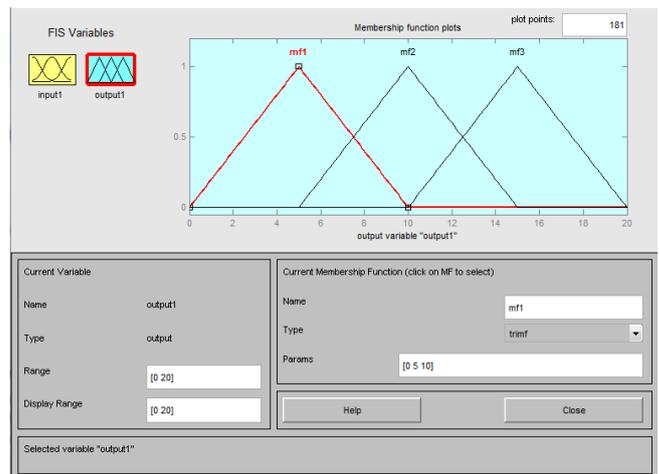

Figure 3: Fuzzy Logic Output Membership

Fuzzy Logic Rules:
If Input = 2 transactions then output = 5
If Input = 6 transactions then output = 10
If input = 10 transactions then output = 15

Rather than classifying the use cases into three classes (simple, average and complex) as in Karners's work, the use cases will be classified into ten categories according to the number of transactions per use case. Since the main goal of our approach is to enhance the current model proposed by Karner and not to completely modifying it, we assume that the largest use case contains ten transactions. We also assume that the complexity factor of the largest use case is fifteen. Table 3 presents a comparison between the original work (Karner's method) and the proposed fuzzy logic approach. The table shows that in the proposed approach, the weights of the use cases are gradually increasing as opposed to the abrupt increase in Karner's method.

Table 3: Adjusted Weight

| Use case contains | Karner's weight | Adjusted weight |
|---|---|---|
| 1 transaction | 5 | 5 |
| 2 transactions | 5 | 5 |
| 3 transactions | 5 | 6.45 |
| 4 transactions | 10 | 7.5 |
| 5 transactions | 10 | 8.55 |
| 6 transactions | 10 | 10 |
| 7 transactions | 10 | 11.4 |
| 8 transactions | 15 | 12.5 |
| 9 transactions | 15 | 13.6 |
| 10 transactions | 15 | 15 |

**PROPOSED MODEL USING NEURAL NETWORK APPROACH**

In this stage, a neural network approach is used to map the input vectors (use cases and actors) to an output vector (UUCP) as shown in Figure 4. Since the nature of the problem is non linear, Multi Layer Perceptron with one hidden layer was used to simulate the problem. There are thirteen input vectors. These include ten vectors that represent the use cases and three vectors that represent the actors.

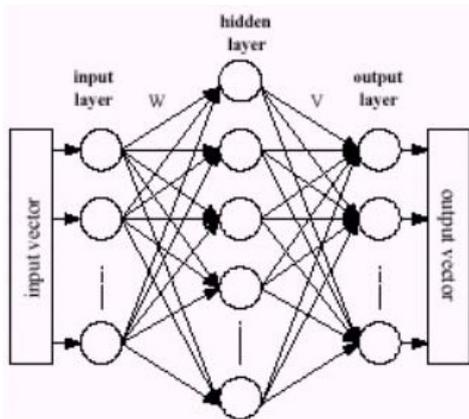

Figure 4: Multi Layer Perceptron

The training algorithm used was Levenberg-Marquardt backpropagation (trainlm). Several experiments were conducted to determine the number of neurons in the hidden layer. As a rule of thumb, the number of neurons in the hidden layer must be greater than the number of neurons of the input layer. However, there are no standard rules to determine the number of neurons in the hidden layer other than trial and error [21]. Twelve experiments were performed. The number of neurons was set between fourteen and twenty five. The best results were obtained when the number of neurons in the hidden layer was twenty. Seven projects were used in training the neural network and thirteen projects were used for testing and validation. The next section demonstrates the results of applying the neural network approach.

**EVALUATION**

The evaluation of this work was conducted on twenty different projects. There is no standard and known conversion between the size in UCP and the size in function points or SLOC. Since some information about the complexity of the projects and the team experience is known about each project, the Technical Factor (TF) and the Environmental Factor (EF) were calculated. Karner suggested that the effort required to develop one UCP is twenty person hours. This method had been criticized by many researchers. Schneider et al. [22] refined Karner's method in determining the effort from UCP. Schneider suggested counting the number of factor ratings of F1-F6 in Table 2 (Technical Factors) that are below three and the number of factor ratings of F7-F8 that are above three. If the total is three or less, then twenty person hours per UCP should be used. If the total is three or four, twenty eight person hours per UCP should be used. If the total is five or more, then the project team should be reconstructed so that the numbers fall at least below five. A value of five indicates that this project is at significant risk of failure with this team. In this paper, Schneider's method has been used to calculate the size of the projects in UCP from the effort. Equation 2 is used to determine the size of each project in UUCP. To distinguish between the results in the proposed fuzzy logic and neural network approaches, the evaluation of each approach was done separately. Furthermore, to determine the effect of the extension part of the use case scenario on size, two different experiments were conducted.

**Evaluation of the Fuzzy Logic Approach**

Karner ignored the "extend" and "include" use cases when applying the UCP model, however we believe that the "extend" and "include" use cases of the use case model should be considered when estimating the size of software. The evaluation of the fuzzy logic approach was conducted in three different stages. First, the evaluation was done on seven projects. The use case models of these projects contain no or very few "extend" and "include" use cases. In the second stage, the evaluation was done on five projects. The use case models of these projects contain a fair number of "extend" and "include" use cases. In this stage, the number of "extend" and "include" use cases in each project is between 15% to 25% of the number of total use cases. Finally, in the third stage, eight projects were chosen for evaluation. In these projects, the number of the "extend" and "include" is more than 25% of the number of total use cases. In each stage, the error (MER, and MRE) of each project was calculated between the original size in UUCP and each of Karner's method and the proposed fuzzy logic approach. At the end of each stage, the error was presented as MMRE, MMER and mean error with standard deviation. Table 4 shows a comparison between the Karner's model and the proposed fuzzy logic approach.

Table 4: Comparison between Karner's and the Proposed Models

| Project | Actual Size UUCP | Karner's Estimation | PropoSed Model (Fuzzy) | MRE Karner | MRE Fuzzy Logic | MER Karner | MER Fuzzy Logic | Error Karner (Karner–Actual) | Error Fuzzy (Fuzzy–Actual) |
|---|---|---|---|---|---|---|---|---|---|
| Project 1 | 72.44 | 128.96 | 104.98 | 0.78 | 0.45 | 0.44 | 0.31 | 56.52 | 32.54 |
| Project 2 | 74.33 | 128.54 | 108.65 | 0.73 | 0.46 | 0.42 | 0.32 | 54.21 | 34.32 |
| Project 3 | 55.50 | 51.00 | 48.70 | 0.08 | 0.12 | 0.09 | 0.14 | -4.50 | -6.80 |
| Project 4 | 68.00 | 108.50 | 92.40 | 0.60 | 0.36 | 0.37 | 0.26 | 40.50 | 24.40 |
| Project 5 | 48.75 | 74.25 | 61.25 | 0.52 | 0.26 | 0.34 | 0.20 | 25.50 | 12.50 |
| Project 6 | 94.50 | 168.75 | 144.00 | 0.79 | 0.52 | 0.44 | 0.34 | 74.25 | 49.50 |
| Project 7 | 72.50 | 108.41 | 92.44 | 0.50 | 0.28 | 0.33 | 0.22 | 35.91 | 19.94 |
| | | | | | | | | | |
| Mean | | | | 0.57 | 0.35 | 0.35 | 0.26 | 40.34 | 23.77 |
| Standard Dev | | | | | | | | 25.33 | 17 |
| Improvement | | | | +22% | | +9% | | | |
| | | | | | | | | | |
| Project 8 | 96.80 | 81.05 | 74.82 | 0.16 | 0.23 | 0.19 | 0.29 | -15.75 | -21.98 |
| Project 9 | 79.80 | 98.67 | 84.54 | 0.24 | 0.06 | 0.19 | 0.06 | 18.87 | 4.74 |
| Project 10 | 91.50 | 118.45 | 109.75 | 0.29 | 0.20 | 0.23 | 0.17 | 26.95 | 18.25 |
| Project 11 | 86.58 | 63.21 | 65.12 | 0.27 | 0.25 | 0.37 | 0.33 | -23.37 | -21.46 |
| Project 12 | 188.64 | 132.54 | 128.67 | 0.30 | 0.32 | 0.42 | 0.47 | -56.10 | -59.97 |
| | | | | | | | | | |
| Mean | | | | 0.25 | 0.21 | 0.28 | 0.26 | -9.88 | -16.08 |
| Standard Deviation | | | | | | | | 33.67 | 30.01 |
| Improvement | | | | +4% | | +2% | | | |
| | | | | | | | | | |
| Project 13 | 94.36 | 54.88 | 48.44 | 0.42 | 0.49 | 0.72 | 0.95 | -39.48 | -45.92 |
| Project 14 | 87.44 | 52.87 | 46.55 | 0.40 | 0.47 | 0.65 | 0.88 | -34.57 | -40.89 |
| Project 15 | 111.50 | 75.84 | 62.54 | 0.32 | 0.44 | 0.47 | 0.78 | -35.66 | -48.96 |
| Project 16 | 119.88 | 67.84 | 72.59 | 0.43 | 0.39 | 0.77 | 0.65 | -52.04 | -47.29 |
| Project 17 | 144.60 | 86.17 | 74.85 | 0.40 | 0.48 | 0.68 | 0.93 | -58.43 | -69.75 |
| Project 18 | 102.87 | 82.40 | 72.88 | 0.20 | 0.29 | 0.25 | 0.41 | -20.47 | -29.99 |
| Project 19 | 124.60 | 64.21 | 52.62 | 0.48 | 0.58 | 0.94 | 1.37 | -60.39 | -71.98 |
| Project 20 | 168.65 | 72.89 | 61.25 | 0.57 | 0.64 | 1.31 | 1.75 | -95.76 | -107.40 |
| | | | | | | | | | |
| Mean | | | | 0.40 | 0.47 | 0.72 | 0.97 | -49.60 | -57.77 |
| Standard Deviation | | | | | | | | 23.00 | 24.47 |
| Improvement | | | | -7% | | -25% | | | |

In the first stage, there is 22% improvement in MMRE and 9% improvement in MMER by applying the proposed fuzzy logic approach. According to equation 13, the mean error with standard deviation of Karner's method can be expressed as 40.34 ±25.33. However, for the fuzzy logic approach, the mean error with standard deviation is 23.77 ±17. In the second stage, there is slim improvement in the proposed method. The MMRE is enhanced by 4% and the MMER is only enhanced by 2%. In the third stage, the new approach has a negative impact and Karner's estimation provided better results. Section 6 will address this change in the results.

**Evaluation of the Neural Network Approach**

Seven random projects were selected to train the neural network presented in section 4. The neural model was tested and evaluated over thirteen projects. Good results were obtained in the training process. The mean error was 0.0215, and the standard deviation was 0.0616. Table 5 presents the results of the neural network approach.

Table 5: Comparison between Karner's and Neural Network Approach

|  | MRE (Karner) | MRE (Neural Network) | MER (Karner) | MER (Neural Network) | Error (Karner) | Error (Neural Network) |
|---|---|---|---|---|---|---|
| Mean | 0.44 | 0.79 | 0.51 | 0.31 | 36.15 | 49.45 |
| Standard Deviation |  |  |  |  | 23.66 | 33.89 |
| Improvement | -35% |  | +20% |  |  |  |

The results show that an improvement of 20% in the MMER was obtained. Table 5 also shows that the neural network approach had adverse results in the MMRE and in the mean error with standard deviation. Section 6 will discuss the results of the neural network approach.

**Effect of the Extension Part in the Use Case Scenario on Size Estimation**

According to Karner's model, the transactions in the extensions are counted the same way as in the main scenario. Two experiments were performed on two projects (project 3 and project 4) to learn the effect of the extension part on size estimation. There were two reasons for choosing these projects. First, the number of "extend" and "include" use cases in these two projects is about 5% of the number of total use cases in the use case diagram. This is important to put the problem of counting the "extend" and "include" use cases aside while working with extensions. Secondly, we are very familiar with these projects. Surprisingly, the MMRE and the MMER of both Karner and the fuzzy logic approach had improved when the extension part of the scenario was ignored. This concluded that in the first stage of projects (project 1 to project 7) where the number of "extend" and "include" use cases is very low, one of the reasons behind the overestimation in both Karner's and the fuzzy logic approach was counting the transactions in the extension part the same way as in the success scenario. For instance, in the projects where the number of transactions in the extensions is approximate to the number of transactions in the success scenario (like the scenario proposed in section 2.1), counting the transactions of the extensions in the same way as in the success scenario might lead to overestimation in the software effort by 30% to 50%.

**DISCUSSION**

Upon conducting experiments in this paper, some important points are noted. These include:
- In about 80% of the projects, the average size of the projects using the fuzzy logic approach was less than the average size of the projects using Karner's approach. This is because the fuzzy logic approach provided a gradual and smooth increase of the complexity weights of the use cases as opposed to the abrupt change in Karner's model.
- Karner did not consider the "include" and the "extend" use cases when counting the transactions in each use case, however the number of "extend" and "include" use cases has an impact on estimation and should be considered. However, more research is required to compare the effort needed to develop the "extend" and "include" use cases with the effort needed to develop the main use cases. In a nut shell, counting the "extend" and "include" use cases might differ from counting the main use cases. Furthermore, the experiments show that Karner's model leads to overestimation when there are no "extend" or "include" use cases. On the other hand, Karner's method gives better results when there is a fair number of "extend" and "include" use cases. It might be concluded that Karner made a rough estimation when he assigned the complexity weights by indirectly including the "extend" and "include" use cases.
- Regarding the extensions in the use case scenario, the transactions in the extension part should be considered, but they should be counted in a different way than in the success scenario. For instance, in the scenario proposed in section 2.1, the number of transactions in the extension part is larger than the number of transactions in the success scenario. Nonetheless, the effort required to develop the extension part might be about 30% of the effort required to develop the success scenario.
- According to Karner, the actor that interacts with five use cases has the same value as if it interacts with one use case. In practice, this might be incorrect. However, since the weight of actors is very low in comparison with use cases, the error is negligible, especially in large projects.
- The results of the fuzzy logic approach were better than the Karner's model in the first two stages (see Table 4). However, the fuzzy logic approach could not beat Karner's model in stage three. The main reason is that the average size of these projects is large and an assumption was made in Section 3 to set the complexity weight of the largest use case to fifteen as Karner proposed. Had the complexity weight of the largest use case been greater than fifteen, the fuzzy logic approach would have given better results.
- The results of the neural network were good in the MMER and not as favourable in the MMRE. This is because more projects are required for training and testing. Moreover, in some situations, the MMRE and the MMER work against each other. This means that improving the MMRE might worsen the MMER and vice versa. This is because the denominator in the MRE is the actual value, however the denominator in the MER is the estimated value.

## THREATS TO VALIDITY

In these experiments, threats to validity can be summarized as follows:

- In the neural network approach, promising results were obtained in the training phase, however this model was not effective in the testing phase when using the MMER criteria. The main reason of this is the lack of projects. The industrial projects that are available for evaluation are scarce. This is because industrial firms are not ready to divulge the UML diagrams of their projects.
- Most of the projects used were educational projects. Some students may not follow the steps of the software development life cycle effectively. Moreover, the quality of some projects might be poor and if the same projects are developed in industry, the actual size might be much more than the size obtained by students.
- There were difficulties in calculating the actual size in UCP or UUCP. Since there is no conversion metrics between UCP and other size metrics, the size in UCP was obtained from the effort, and then equation 2 was used to obtain the size in UUCP. Although Schneider's method (Karner's refined method) was used to calculate the size in UCP, this method might not be as accurate as other sophisticated cost estimation models such as SEER-SEM.
- The use case points model mainly depends on the use case diagram. If the use case diagram was not properly designed, a huge error could be incurred.

## CONCLUSION AND FUTURE WORK

The use case points model is one of the cost estimation models that has been widely used because it is simple, fast, accurate to a certain degree and can be automated. The use case points model is based on the number and the complexity of the use cases as well as the actors. The original model suggested three degrees of complexity to the use cases and there is no graduation among the complexity weights of the use cases. This paper presented the disadvantages of the current model and proposed an enhancement to this model using fuzzy logic and neural network. The fuzzy logic approach presents ten degrees of complexity of the use cases. Moreover, this approach provides graduation among the complexity weight. The neural network approach was used as a black box to map the input vectors of the use case model to software size. The results showed that the UCP software estimation can be improved up to 22% in some projects.

Future work will focus on revamping the use case model. First, the largest use case should contain at least twenty transactions as opposed to eight transactions as in Karner's model. Secondly, the complexity weight of the use cases will be calibrated using the neuro-fuzzy approach. Thirdly, "extend" and "include" use cases should be considered when estimating the software size. Finally, the future work will focus on how the "extend" and "include" use cases as well as the transactions in the extension part should be counted.